\documentclass[a4paper,11pt]{article}
\usepackage{jheppub} % for details on the use of the package, please see the JINST-author-manual
\usepackage[dvipsnames]{xcolor}
\usepackage{verbatim}
\usepackage[bottom]{footmisc}

\usepackage{tikz}
\usetikzlibrary{matrix}
\usepackage{filecontents}

\newcommand{\pars}[1]{\left(#1\right)}
\newcommand{\spars}[1]{\left[#1\right]}

\definecolor{CByellow}{RGB}{255, 176, 0}
\definecolor{CBmagenta}{RGB}{220, 38, 127}
\definecolor{CBviolet}{RGB}{120, 94, 240}

\arxivnumber{2406.16124} % if you have one

\title{\boldmath Geometrisation of Ohm's reciprocity relation in a holographic plasma}

\author{Giorgio Frangi}
\affiliation{Higgs Centre for Theoretical Physics, The University of Edinburgh, \\
Peter Guthrie Tait Road, Edinburgh EH9 3FD, Scotland}

\emailAdd{giorgio.frangi@ed.ac.uk}

\abstract{It has been recently pointed out that the familiar reciprocity relation between the conductivity $\sigma$ and resistivity $\rho$, which I refer to as \textit{Ohm's reciprocity relation}, should not be expected to hold in all possible settings, but is rather a property that may (or may not) emerge as a consequence of specific features, or in certain limits of interest, of a given theory. In this work I prove an analogous statement: $\rho= \sigma^{-1}$, across two different classes of holographic theories related by a generalisation of the electric-magnetic duality in the $D=4+1$ bulk. In terms of the dual hydrodynamic theories, this statement is shown to imply the suppression of any contributions to the transport coefficients from dynamical electromagnetic fields, present in only one of the two theories. This makes the two theories, as far as late-time linear electric transport is concerned, equivalent. I then confirm these findings by considering one specific model and run numerical simulations in different settings. }

\begin{document}

\maketitle
\flushbottom

\section{Introduction}

Electric conduction in a vast and very diverse array of materials is described by Ohm's law, which prescribes that, under very few assumptions, applying a voltage $V$ to any two points of a piece of a conducting material results in a linearly proportional electric current $I$ flowing between them: $V=R \cdot I$ \cite{kittel}. The coefficient $R$ is the \textit{resistance} of the conductor, which depends on both its shape and the intrinsic and thermodynamic properties of the material. To isolate the latter, it is possible to rewrite Ohm's law in terms of locally defined transport coefficients, named the \textit{conductivity} $\sigma$ and the \textit{resistivity} $\rho$, which express a linear relation between the current density $\mathbf j$ and the electric field $\mathbf E$ measured at a given point in space. In simple phenomenological models -- for example in the Drude model --, the two of them are effectively interchangeable, as they are taken to be reciprocal: $\rho = \sigma^{-1}$. I will refer to this identity as \textit{Ohm's reciprocity relation}. Importantly, such a relationship is also featured in effective descriptions of magnetohydrodynamics (MHD) in terms of a derivative expansion \cite{Hernandez_2017}.

At a more fundamental level, transport coefficients such as $\sigma$ and $\rho$ can be related to retarded correlators of an underlying quantum field theory (QFT) via the machinery of linear response theory. In this context, they are defined by appropriate Kubo formulae \cite{Kubo:original}:
\begin{equation}
    \label{eq:kubos}
    \sigma (\omega) = \lim_{\mathbf k \to 0} \frac{G^{jj}_\text{ret}(\omega,\mathbf k)}{i \omega}, \quad \rho(\omega) = \lim_{\mathbf k \to 0} \frac{G^{EE}_\text{ret}(\omega,\mathbf k)}{i \omega},
\end{equation}
where $G^{XX}_\text{ret}$ denotes the real-time retarded two-point function of the operator $\hat X$. While the former is a well known result that has been present in the literature for a long time -- see for example  \cite{AMYpaper,Kovtun:2012rj,KapustaGale} --, the expression for $\rho$ is much more recent, and was originally reported in \cite{GHI_2formmhd} in an equivalent form involving the retarded 2-point function of a conserved magnetic 2-form symmetry $\mathcal J^{\mu\nu}$:%\footnote{The equivalence of the two expressions for $\rho$ has been discussed in \cite{Hernandez_2017}.}
\begin{equation}
    \label{eq:kubos_app}
    \rho(\omega) = \lim_{\mathbf k \to 0} \frac{G^{\mathcal J \mathcal J}_\text{ret}(\omega,\mathbf k)}{i \omega}.
\end{equation}
Notably, for the resistivity $\rho$ to be well defined one requires electromagnetic fields to be dynamical in the theory under consideration; no corresponding requirement exists for the conductivity. The microscopic perspective encoded in the Kubo formulae \eqref{eq:kubos} and \eqref{eq:kubos_app} then clarifies that, given such a theory, there is no obvious relation between $\sigma$ and $\rho$, so that the reciprocity relation becomes, in effect, a law of nature to be proven. One way to test it would then be to explicitly calculate both transport coefficients using either \eqref{eq:kubos} or \eqref{eq:kubos_app}, and compare the two results. A first study in this sense has been \cite{toapp}, which concluded that, in absence of a finite conserved charge density, $\rho = \sigma^{-1}$ can be shown to be an emergent property of $T \neq 0$ states either in the DC limit ($\omega\to0$) or, in very special theories in which higher order diagrammatic corrections are suppressed, even at finite frequency. Remarkably, in the latter case the resistivity has been seen to be reciprocal to the conductivity calculated in absence of dynamical electromagnetic fields.

The primary goal of this work is to investigate these claim in a more general setting, by focusing on the relatively simple case of states with a finite background charge density in $d=3+1$ dimensions. To this purpose, I will restrict to theories which possess a classical gravitational dual \cite{Maldacena:1997re}. This will avoid technical complications that arise in the calculation of retarded thermal correlators in the IR regime when adopting other techniques, such as the breakdown of perturbation theory in ordinary thermal QFT -- see \cite{Jeon:long,Jeon:Yaffe} --, or having to solve a quantum Boltzmann equation while working with a large $N$ expansion \cite{Damle_1997,Sachdev_1998}. In case a classical gravitational dual is available, these correlators can be computed from the bulk physics, following the relatively simple prescription outlined in \cite{SS_minkPrescr}.  Building on this result, many subsequent studies have then established a dictionary to study thermodynamic and transport properties of phases of matter with realistic features, see for example \cite{Hall_HK,Hartnoll-lectures,aw_axions,DHK_nMB,hartnoll2018holographic,axion_review}. 

One important entry of this dictionary is that conserved $n$-form currents on the boundary correspond to rank-$n$ gauge fields in the bulk. Thus, holographic conductivities can be calculated from gravitational theories comprising an electromagnetic sector (such as AdS-Einstein-Maxwell, see for example \cite{hartnoll2018holographic}), whereas resistivities require the presence of a dynamical 2-form gauge field, as proposed in \cite{GrozPoov_holoHF,HofIqb_holoHF}. This poses a problem to the announced goal of this work, as the operator content of a theory in which computing $\sigma$ is possible may differ from one allowing the calculation of $\rho$; furthermore, the fact that $\rho$ has be calculated in a theory with dynamical electromagnetism requires the adoption of mixed boundary conditions \cite{dynboundary}, rather than the Dirichlet ones in adoption for $\sigma$. One could then wonder whether such a comparison would be meaningful in the first place.

In general, one does not expect this to be the case. Nevertheless, restricting to the special case of gravitational backgrounds encoding a homogeneous charge density, it is possible to make a precise comparison, which is rooted in a duality transformation generalising the electric-magnetic duality of free Maxwell theory in $D=3+1$ bulk dimensions to one dimension more, which crucially maps Dirichlet boundary conditions into mixed ones in the process \cite{DeWolfeHiggin}. This requires to make the notion of Ohm's reciprocity looser, by allowing the comparison between two different hydrodynamic theories defined on the same thermal ensemble. This is the subject of this work, and its details are elaborated for a specific model in Section \ref{sec:bcg}, in which I also discuss limitations to this procedure, and its interpretation from the viewpoint of hydrodynamics.

Four further sections follow. In Section \ref{sec:ren} I consider the problem of holographic renormalisation and elaborate on the scale-dependence of the holographic Kubo formulae for $\sigma$ and $\rho$ in $d=3+1$ boundary dimensions. These results are key in proving Ohm's reciprocity relation in the subsequent Section \ref{sec:analytic_proof}, where I show that this identity is encoded in special properties of the background bulk metric, thus reinforcing the importance of the restrictions mentioned earlier. To further substantiate my claim, in Section \ref{sec:tc} I reprise the backgrounds studied in Section \ref{sec:bcg} and proceed with a numerical calculation of the transport coefficients, showing the reciprocity relation to hold even at finite frequency. I will then conclude in Section \ref{sec:disc} by outlining the boundaries of validity of the result and discuss a few possible phenomenological application of the result, together with possible future developments.

\section{Dualisation map and background}
\label{sec:bcg}

In holographic calculations of response functions, all dynamical fields are separated into background values and fluctuations, with the former encoding the thermodynamic properties of the boundary states upon which the latter are calculated. The aim of this section is to review how to select thermal states with a finite conserved charge density for the calculation of holographic conductivities, how to achieve the same for resistivities, and to show that the duality mentioned in the introduction is, for this special case, state-preserving -- thus making a comparison between quantities calculated in the two theories meaningful. While I will do this for a specific model, the procedure can be easily repeated for a number of more complicated ones, with exceptions discussed at the end of the section. From now on, I specialise to homogeneous, isotropic states in $d=3+1$ boundary dimensions -- unless otherwise stated.

\medskip

\paragraph{Conductivity}

A simple model to calculate the conductivity of a charged plasma dissipating momentum has been given by Andrade and Withers \cite{aw_axions}. They consider the gravitational action:
\begin{equation}
    \label{eq:aw_action}
    S_\sigma = \int d^5 x \sqrt{-g} \left( R - 2 \Lambda - \frac{1}{4e^2} F_{ab}F^{ab} - \frac{1}{2} \sum_I \partial_a \psi^I \partial^a \psi^I  \right) + S^\text{bdy}_\sigma,
\end{equation}
where $\Lambda = - 6/L_\text{AdS}^2$ is the (negative) cosmological constant, $F = dA$ is the tensor strength associated to an abelian one-form gauge connection and $\psi^I$ is a multiplet of scalar fields (axions) with the index $I$ running on the spatial coordinates of the boundary QFT: $x,y$ and $z$. The radial coordinate $r$ is picked so that the AdS boundary sits at $r\to\infty$. $S_\text{bdy}$ is a local action defined on the AdS boundary including the Gibbons-Hawking-York term and possibly counterterms -- which I discuss later. For simplicity, from now on I set both the gauge coupling $e$ and $L_\text{AdS}$ to unity.

The equations of motion following from \eqref{eq:aw_action} are:
\begin{equation}
    \label{eq:aw_covariant_eoms}
    \begin{gathered}
        \Box \psi^I = 0, \\
        \nabla_a F^{ab} = 0, \\
        R_{ab} + 4 g_{ab} - \frac{1}{4} \left( 2 F_{ac}{F_b}^c - \frac{1}{3} g_{ab} F^2 \right) - \frac{1}{2}\partial_a \psi^I \partial_b \psi^I = 0.
    \end{gathered}
\end{equation}
As previously stated, I will separate each field into background and fluctuation. Formally:
\begin{equation}
    \psi^I \to \psi^I + \delta \psi^I, \quad A_a \to A_a + \delta A_a, \quad g_{ab} \to g_{ab} + \delta g_{ab}.
\end{equation}
In the immediate following I deal with the background part only; fluctuations will be important when considering transport in later sections. In order to restrict to homogeneous, isotropic, static states with no background flow, one picks the ansatz:
\begin{equation}
    \label{eq:aw_bcg_ans}
    \begin{gathered}
        \psi^I = \alpha \, x^I, \quad A = A_t (r) \, dt, \\ ds^2 = -f(r) dt^2 + \frac{dr^2}{f(r)} + r^2 dx^I dx^I.
    \end{gathered}
\end{equation}
The metric coefficient $f$ is taken to have a simple zero at a finite radius $r_h$: $f(r_h)=0$. This inserts a black brane event horizon in the bulk, which places the boundary theory in a thermal bath. Additionally, $A_t(r_h)=0$ to ensure the gauge field to be regular on the horizon. The linear profile of the axions spoils neither homogeneity nor isotropy as only their derivatives appear in the action \eqref{eq:aw_action}.

The expressions \eqref{eq:aw_bcg_ans} can be plugged into the covariant equations \eqref{eq:aw_covariant_eoms} to give two ODEs for $A_t$ and $f$. These can in turn be solved with the aforementioned boundary conditions on the horizon $r_h$ and the additional requirement that the time component of the gauge field asymptotes to a constant value $\mu$, which can be interpreted as a chemical potential in the boundary theory. This leads to the background functions:
\begin{gather}
    \label{eq:At_bcg}
    A_t(r) = \mu \pars{1 - \frac{r_h^2}{r^2}}, \\[4pt]
    \label{eq:f_bcg}
    f(r) = \pars{1 - \frac{r_h^2}{r^2}}\pars{r^2 + r_h^2 - \frac{\alpha^2}{4} - \frac{\mu^2 r_h^2}{3 r^2}}.
\end{gather}
The chemical potential $\mu$ can be expressed in terms of the charge density $n$ by noticing that the $r\gg1$ expansion of $A_t$ is, according to the holographic dictionary:
\begin{equation}
    A_t(r) \simeq \mu - \frac{ n }{2 r^2},
\end{equation}
so that: $n = 2 r_h^2 \mu$. The two variables can then be used interchangeably. It is nevertheless important to stress that fixing one in terms of the other is equivalent to a choice of thermodynamic ensemble. In particular, when adopting Dirichlet boundary conditions -- see also the later sections --, $n$ should be interpreted as the expectation value of the charge density operator $j^0$ over the ensemble specified by the externally fixed chemical potential $\mu$:
\begin{equation}
	\label{eq:nvev}
	n = \langle \hat j^0 \rangle_\mu. 
\end{equation}
In this sense, fixing $n$ corresponds to perform a Legendre transformation on the boundary; well-posedness of this operation is ensured by the invertibility of the function $n(\mu)$.

Each bulk background configuration for the model \eqref{eq:aw_action} is uniquely identified by 3 dimensionful parameters: $\alpha$, $n$ (or $\mu$) and the temperature $T$, which is equal to:
\begin{equation}
    \label{eq:T_H}
    T = \frac{1}{\pi r_h} \pars{r_h^2 - \frac{\alpha^2}{8} - \frac{n^2}{24 r_h^4}}.
\end{equation}
As the boundary theory is scale invariant, its states will be labeled by two dimensionless ratios. For the rest of this work, I will take them to be: $\alpha/T$ and $n/T^3$. \medskip

\paragraph{Resistivity}

In order to calculate a holographic resistivity from the Kubo formula \eqref{eq:kubos_app} one needs to introduce a bulk gauge field dual to a 2-form conserved current, as outlined in \cite{GrozPoov_holoHF,HofIqb_holoHF}. With the addition of axions, a minimal action of this kind is:
\begin{equation}
    \label{eq:hf_action}
    S_\rho = \int d^5 x \sqrt{-g} \left( R - 2 \Lambda - \frac{1}{12 \tilde e^2} H_{abc}H^{abc} - \frac{1}{2} \sum_I \partial_a \psi^I \partial^a \psi^I  \right) + S^\text{bdy}_r.
\end{equation}
Here, $H=dB$ with $B$ a 2-form gauge field. Its gauge freedom is $B \to B + d\Lambda$, with $\Lambda$ a generic 1-form. For simplicity, I will take $\tilde e = 1$. It is then straightforward to derive the covariant equations of motion:
\begin{equation}
    \label{eq:hf_covariant_eoms}
    \begin{gathered}
        \Box \psi^I = 0, \\
        \partial_a (\sqrt{-g} H^{abc} ) = 0, \\
        R_{ab} + 4 g_{ab} - \frac{1}{4} \left(H_{acd}{H_b}^{cd} - \frac{2}{9} g_{ab} H^2 \right) - \frac{1}{2}\partial_a \psi^I \partial_b \psi^I = 0.
    \end{gathered}
\end{equation}
Again, fields can be separated into background and fluctuations:
\begin{equation}
    \psi^I \to \psi^I + \delta \psi^I, \quad B_{ab} \to B_{ab} + \delta B_{ab}, \quad g_{ab} \to g_{ab} + \delta g_{ab}.
\end{equation}

From here, one could conduct the same analysis outlined above, with the explicit aim to select an equivalent thermodynamic state -- so to make any further comparison meaningful. This time, however, the charge density will appear as a source, rather than an expectation value. This can be understood from the fact that the conserved magnetic 2-form current that is dual to $B_{ab}$ is the dual electromagnetic field strength:
\begin{equation}
	\mathcal J = *_4 F,
\end{equation}
where, in this instance only, $F$ refers to the electromagnetic field strength on the boundary. The conservation law for the higher-form current $\mathcal J$ is none other than the Bianchi identity: $dF=0$, i.e. the constraints among Maxwell's equation. In terms of $\mathcal J$, Gauss' law reads:
\begin{equation}
	*_3 d_3 \mathcal J = n,
\end{equation}
where the subscripts denote that both the exterior derivative and the Hodge dualisation should be taken only on the spatial coordinates of the boundary. Using the holographic dictionary, this constraint is satisfied by:
\begin{equation}
    \label{eq:H_bcg}
    H = n \, dx \wedge dy \wedge dz,
\end{equation}
which solves the equations of motion \eqref{eq:hf_covariant_eoms}, and can be taken to be the desired background solution. Notably, \eqref{eq:H_bcg} can also be obtained by considering model \eqref{eq:aw_action} and applying the transformation \cite{DeWolfeHiggin}:
\begin{equation}
    \label{eq:duality5}
    H = *_5 F,
\end{equation}
to the solution \eqref{eq:aw_bcg_ans}, where $*_5$ is the Hodge star in the bulk. In previous studies, this map has been applied to relate fluctuations of $A_a$ and $B_{ab}$ on top of neutral backgrounds, but because it relates the two models \eqref{eq:aw_action} and \eqref{eq:hf_action} at the level of the full action:
\begin{equation}
    F \wedge *_5 F = - *_5 H \wedge H = H \wedge *_5 H,
\end{equation}
one is led to expect that such relationship should extend to backgrounds as well.

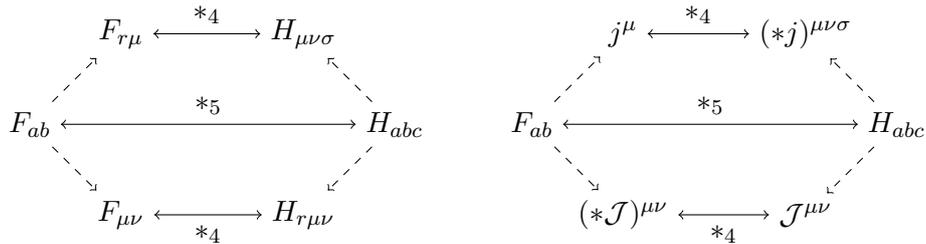
\begin{figure}
    \centering
    \begin{tikzpicture}[scale=1.2]
        \node (F5) at (0,1) {$F_{ab}$};
        \node (F4v) at (1,2) {$F_{r\mu}$};
        \node (F4t) at (1,0) {$F_{\mu\nu}$};
        \node (H4v) at (3,2) {$H_{\mu\nu\sigma}$};
        \node (H4t) at (3,0) {$H_{r\mu\nu}$};
        \node (H5) at (4,1) {$H_{abc}$};
        \draw
            (F5) edge[->] [dashed] (F4v)
            (F5) edge[->] [dashed] (F4t)
            (H5) edge[->] [dashed] (H4v)
            (H5) edge[->] [dashed] (H4t)
            (F4v) edge[<->] node[above] {$*_4$} (H4v)
            (F4t) edge[<->] node[below] {$*_4$} (H4t)
            (F5) edge[<->] node[above] {$*_5$}(H5);
        \end{tikzpicture} \qquad
        \begin{tikzpicture}[scale=1.2]
        \node (F5) at (0,1) {$F_{ab}$};
        \node (F4v) at (1,2) {$j^{\mu}$};
        \node (F4t) at (1,0) {$(*\mathcal J)^{\mu\nu}$};
        \node (H4v) at (3,2) {$(*j)^{\mu\nu\sigma}$};
        \node (H4t) at (3,0) {$\mathcal J^{\mu\nu}$};
        \node (H5) at (4,1) {$H_{abc}$};
        \draw
            (F5) edge[->] [dashed] (F4v)
            (F5) edge[->] [dashed] (F4t)
            (H5) edge[->] [dashed] (H4v)
            (H5) edge[->] [dashed] (H4t)
            (F4v) edge[<->] node[above] {$*_4$} (H4v)
            (F4t) edge[<->] node[below] {$*_4$} (H4t)
            (F5) edge[<->] node[above] {$*_5$}(H5);
        \end{tikzpicture}
    \caption{\textbf{Left:} effect of the $5d$ Hodge dualisation on the $4d$ forms induced on a radially constant slice. \textbf{Right:} schematic effect on the corresponding conserved currents. By definition: $\mathcal J = *F$. Dashed arrows denote induction operations, whereas solid ones Hodge dualisations.}
    \label{fig:stars_diagram}
\end{figure}

There is an elegant interpretation of this: given a 2-form (such as $F_{ab}$) in 5$d$ and a $4d$ brane at constant $r$, the former induces a 1- ($F_{r\mu}$) and a 2-form ($F_{\mu\nu}$) on the latter. Likewise, a 3-form (such as $H_{abc}$) induces a 2- ($H_{r\mu\nu}$) and a 3-form ($H_{\mu\nu\sigma}$). If the higher-dimensional forms are related by Hodge dualisation in the bulk ($*_5$), the induced fields will likewise be related by Hodge dualisation restricted to the brane ($*_4$), up to multiplicative $r$-dependent factors. This is depicted in the diagram in the left panel of Figure \ref{fig:stars_diagram}.

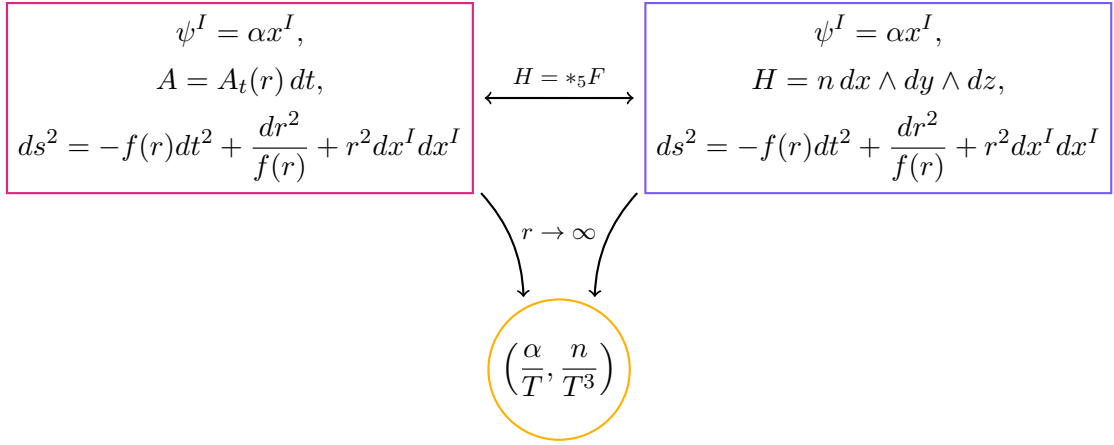
\begin{figure}
    \centering
    \begin{tikzpicture}[scale=1.2]%,every node/.append style={circle, draw=blue!80, inner sep=0pt, minimum size=12pt}]
        \node[circle, draw=CByellow, thick, inner sep=2pt, minimum size=12pt] (B) at (0,0) {$\begin{aligned} \pars{\frac{\alpha}{T},\frac{n}{T^3}} \end{aligned}$};
        \node[thick, draw=CBmagenta, inner sep=4pt, minimum width=18pt,minimum height=8pt] (L) at (-3.5,3) {$\begin{gathered} \psi^I = \alpha x^I, \\ A = A_t(r) \, dt, \\ ds^2 = -f(r)dt^2 + \frac{dr^2}{f(r)} + r^2 dx^I dx^I  \end{gathered}$};
        \node[thick, draw=CBviolet, inner sep=4pt, minimum width=18pt,minimum height=8pt] (R) at (3.5,3) {$\begin{gathered} \psi^I = \alpha x^I, \\ H = n \, dx \wedge dy \wedge dz, \\ ds^2 = -f(r)dt^2 + \frac{dr^2}{f(r)} + r^2 dx^I dx^I  \end{gathered}$};
        \node (label) at (0,1.5) {\footnotesize $r \to \infty$};
        \draw
            (L) edge[<->,thick,shorten >= 4pt,shorten <= 4pt] node[above] {\scalebox{.8}{$H = *_5 F$}} (R)
            (L) edge[->,thick,shorten >= 4pt,shorten <= 4pt] [bend left=20] (B) % node[left]{\scalebox{.8}{$r\to\infty$}} 
            (R) edge[->,thick,shorten >= 4pt,shorten <= 4pt] [bend right=20] (B); % node[right]{\scalebox{.8}{$r\to\infty$}} 
        \end{tikzpicture} 
    \label{fig:dima}
    \caption{The selected backgrounds for $S_\sigma$ (magenta, left) and $S_\rho$ (violet, right) give rise to the same boundary state (yellow, bottom) when related by the duality \eqref{eq:duality5}.}
\end{figure}

More importantly, these quantities can be understood in terms of conserved quantities of the boundary theory and their sources. According to the membrane paradigm \cite{IqbalLiu_MB}, expectation values of a conserved current can be extended into the bulk by considering the conjugate momenta of the dual fields with respect to a radial foliation:
\begin{equation}
    \label{eq:membranes}
    j_\text{mb}^\mu = \frac{\delta S_\sigma}{\delta A_{\mu,r}} = - \sqrt{-g} F^{r\mu}, \quad \mathcal J_\text{mb}^{\mu\nu} = \frac{\delta S_r}{\delta B_{\mu\nu,r}} = - \sqrt{-g} H^{r\mu\nu},
\end{equation}
with $j_\text{mb}^\mu$ and $\mathcal J_\text{mb}^{\mu\nu}$ asymptoting to the expectation value of the corresponding boundary currents at $r\to\infty$. % This identification has the effect of specifying which boundary conditions are appropriate -- in this case, Dirichlet boundary conditions for model \eqref{eq:aw_action} and mixed for \eqref{eq:hf_action}. %The membrane paradigm thus clarifies that $F_{r\mu}$ and $H_{r\mu\nu}$ naturally encode information on the corresponding conserved current, so that the diagram can be reinterpreted as in the right panel of Figure \ref{fig:stars_diagram}. 
Recasting the bulk quantities in terms of boundary ones -- as in the right panel of Figure \ref{fig:stars_diagram} -- one also learns that the duality \eqref{eq:duality5} has the effect of mapping the membrane currents of one model into sources of the other, in an operation that amounts to a Legendre transformation. One can also check that solving the equations of motion \eqref{eq:hf_covariant_eoms} leads to the same metric function $f$ in \eqref{eq:f_bcg} and to the same temperature $T$ \eqref{eq:T_H}, as summarised in the diagram in Figure \ref{fig:dima}.

This leads to the conclusion that the effect of the duality \eqref{eq:duality5} on a background for the model \eqref{eq:aw_action} is to perform a Legendre transform and to select a corresponding ensemble for the model \eqref{eq:hf_action}, characterised by the very same thermodynamic quantities. For this particular class of solutions, then, the two models effectively select the same background thermal state.

\paragraph{The viewpoint from hydrodynamics} The one-to-one correspondence between this type of background solutions of the models \eqref{eq:aw_action} and \eqref{eq:hf_action} informs the strategy the rest of this work follows, and the sense in which quantities calculated from the two models should be compared.

It has long been known that low-frequency, long-wavelength perturbations on top of black brane solutions behave hydrodynamically \cite{Kovtun_holohydro}, and that constitutive relations for energy-momentum fluctuations (and other conserved charges) can be reconstructed by solving equations of motion for bulk fields. In 
this sense, \eqref{eq:aw_action} captures the long-time dynamics of an ordinary conserved current coupled to $T_{\mu\nu}$ and external electromagnetic fields, whereas in \eqref{eq:hf_action} the current is a 2-form, electromagnetic fields are dynamical and an ordinary current appears as an external source. $\sigma$ and $\rho$ are kinetic coefficients of those constitutive equations.

The swap between source and expectation value suggests that the hydrodynamic theories following from \eqref{eq:aw_action} and \eqref{eq:hf_action} are related by a Legendre transformation -- which we have shown to be encoded in the bulk duality \eqref{eq:duality5}. As a consequence, the emergence of Ohm's reciprocity becomes a statement over whether finite temperature and charge density, together with the large $N^2$ expansion entailed by assuming the existence of a classical gravitational dual, suppress effects due the dynamical nature of the electromagnetic fields in the boundary of model \eqref{eq:hf_action}, matching the predictions from a theory with no photon dynamics \eqref{eq:aw_action}.

\paragraph{Limitations to the dualisation procedure} In the example presented above, the duality \eqref{eq:duality5} has been proven to be a boundary ensemble-preserving map by showing that it expresses one bulk action in terms of the other, and that solving the equations of motion on the two sides generates the same background. Such a procedure is not always possible, and it is desirable to have a criterion to establish which gravitational actions that include an electromagnetic sector do have a dual in terms of a 2-form field.

A simple answer comes from considering the action of the dualisation on the equations of motion and Bianchi identity in our specific example:
\begin{equation}
\label{eq:swap}
        \begin{tikzpicture}[scale=1]%,every node/.append style={circle, draw=blue!80, inner sep=0pt, minimum size=12pt}]
        \centering
        \node[] (LU) at (-2.5,.3) {$d*_5 F = 0$};
        \node[] (LD) at (-2.5,-.3) {$d F = 0$};
        \node[] (RU) at (2.5,.3) {$d*_5 H = 0$};
        \node[] (RD) at (2.5,-.3) {$d H = 0$};
        \node[] (C) at (0,.25) {\scriptsize $*_5 F = H$};
        \draw
            (LU) edge[<->,shorten >= 4pt,shorten <= 4pt]  (RD)
            (RU) edge[<->,shorten >= 4pt,shorten <= 4pt] (LD); 
        \end{tikzpicture}
\end{equation}
It is apparent that the action of \eqref{eq:duality5} is to swap the two. Importantly, while Bianchi identity is always of the form $dX=0$ (where $X$ is a tensor strength), the equation of motion is not. Focusing on the 1-form side of the transformation, the presence of matter fields charged under the ordinary $U(1)$ can modify its right hand side. This would lead to a contradiction:
\begin{equation}
    \label{eq:contradiction}
    d *_5 F = *j \xrightarrow{\;*_5 F = H\;} d H = d^2 B = *j \neq 0.
\end{equation}
Models such as the celebrated holographic superconductor presented in \cite{Hartnoll:hsc1,Hartnoll:hsc2} violate this requirement, and thus may not have a dual description in terms of 2-forms. An analogous conclusion is reached if the 1-form action includes a Chern-Simons term: $A\wedge F \wedge F$, as it would if it were obtained through the dimensional reduction of AdS\textsubscript{5}$\times S^5$ \cite{witten1998anti,Chamblin_1999}. 
\begin{comment}
\paragraph{Lower dimensional systems}

It is important to stress that this construction is meaningful only in $5$ bulk dimensions. From the boundary viewpoint, in this case only does the map \eqref{eq:duality5} relate a theory with an ordinary conserved current to another with a 2-form current. This is a consequence of $*_5 F$ being a 3-form.

Another interesting case is the analogue construction in $4$ bulk dimensions, which was thoroughly studied in \cite{MTheory}. Here, the bulk dualisation becomes nothing but the electromagnetic duality, which can be understood, in terms of the boundary theories, as a duality of particle-vortex type. From this picture, a relation similar to the reciprocity law naturally emerges; however, a substantial difference from the case at hand is that the lower dimensionality makes the kinetic term of the photon irrelevant, so that it is removed by flowing to the IR, as argued in the original reference -- see also \cite{Giombi_confCFT}.
\end{comment}

\section{Renormalisation and Kubo formulae}
\label{sec:ren}

In this section, I derive holographic Kubo formulae for the conductivity $\sigma$ and the resistivity $\rho$. Because of divergences in the boundary action(s), these are sensitive to the renormalisation procedure in a precise sense which is reminiscent of the scale dependence in theories of electromagnetism in 4$d$.

The explicit form of the holographic Kubo formula is expressed as customary in Fefferman-Graham (FG) coordinates, which I write in the form adopted in \cite{Taylor_ct}:
\begin{equation}
    \label{FGmetric_x}
    ds^2 = \frac{du^2}{u^2} + \frac{1}{u^2} \gamma_{\mu\nu}(u,x) dx^\mu dx^\nu,
\end{equation}
so that the AdS boundary is located at $u\to0$, and $\gamma_{\mu\nu}$ has the asymptotic expansion:
\begin{equation}
    \gamma_{\mu\nu}(u,x) \simeq \eta_{\mu\nu} + \gamma^{(2)}_{\mu\nu}(x) u^2 + \gamma^{(4)}_{\mu\nu}(x) u^4 + \tilde\gamma_{\mu\nu}(x) u^4 \log{u} + O(u^5 \log{u}).
\end{equation}
The leading coefficient in the expansion is the 4$d$ Minkowski metric $\eta_{\mu\nu}$ in accordance with the spacetime being asymptotically AdS. For reference:
\begin{equation}
    \label{eq:SStoFG}
    u(r) = \exp\spars{- \int \frac{dr}{\sqrt{f(r)}}} \simeq \frac{1}{r} + \frac{\alpha^2}{16 r^3} + O(r^{-5}).
\end{equation}
This can be used to map any asymptotic expansion for $r\to\infty$ into the corresponding one for $u\to0$ (and vice versa). In the latter case, using the given metric expansion into the covariant equations of motions for each of the two gauge fields (\eqref{eq:aw_covariant_eoms} and \eqref{eq:hf_covariant_eoms}) leads to their FG expansions:
\begin{gather}
    \label{eq:A_FG}
    A_\mu(u,x) \simeq A^{(0)}_\mu(x) + A^{(2)}_\mu(x) u^2  + a_\mu(x) u^2 \log u \,  + O(u^3 \log u), \\
    \label{eq:B_FG}
    B_{\mu\nu}(u,x) \simeq B^{(0)}_{\mu\nu}(x) + B^{(1)}_{\mu\nu}(x) \log u + O(u \log u),
\end{gather}
where $a^\mu = -\partial_{\alpha} F^{\alpha\mu}_{(0)}/2$, following the procedure outlined in \cite{Skenderis_notes}; higher order coefficients, if needed (for example in numerical methods, see later), can be determined recursively.  Plugging the expansions respectively into \eqref{eq:aw_action} and \eqref{eq:hf_action}, one obtains the (divergent) boundary actions:
\begin{gather}
    \label{eq:os_s}
     S_{\sigma, \text{on-shell}} \supset  \int_{u = \epsilon} d^4 x \, A^{(0)}_{\mu} \spars{A_{(2)}^{\mu} - \frac{1}{2}\pars{\log{u} + \frac{1}{2}} \partial_\alpha F_{(0)}^{\alpha\mu} } + O(u \log {u}), \\[4pt]
    \label{eq:os_r}
     S_{\rho, \text{on-shell}} \supset \frac{1}{4} \int_{u = \epsilon} d^4 x \, \pars{B^{(0)}_{\mu\nu} + B^{(1)}_{\mu\nu} \log{ u}} B_{(1)}^{\mu\nu} + O({u} \log {u}).
\end{gather}
These should be understood as defined on a cutoff brane (at $u = \epsilon$) close to the AdS boundary, to be regulated via the addition of suitable local counterterms that make the limit $\epsilon\to0$ finite. The gravitational part is the same in both the actions considered and is known to be logarithmically divergent as well \cite{weyl_anomaly}, but this is not relevant to the present discussion. For this choice of background and for the sake of the quantities I will calculate, the introduction of axions does not require the introduction of further counterterms.

The minimal prescription for holographic renormalisation is to simply remove the divergences \cite{Skenderis_notes}, and this is the scheme that, for simplicity, I will adopt. In case such divergences are logarithmic this is nevertheless not a unique choice, as I will argue later.

\paragraph{Minimal renormalisation scheme} To remove the divergences from the electromagnetic sectors of the boundary actions \eqref{eq:os_s} and \eqref{eq:os_r}, one adds the counterterms -- see \cite{Taylor_ct,GrozPoov_holoHF} -- :
\begin{equation}
    S_{\sigma, \text{ct}} = - \frac{1}{4} \log \epsilon \int_{u = \epsilon} d^4 x \, F^{(0)}_{\mu\nu} F_{(0)}^{\mu\nu}, \quad S_{\rho, \text{ct}} = - \frac{1}{4} \log \epsilon \int_{u = \epsilon} d^4 x \, \mathcal H_{\mu\nu} \mathcal H^{\mu\nu},
\end{equation}
where $\mathcal{H}_{bc} = \bar n^a H_{abc}$ with $\bar n$ an (ingoing) normal to the $u=\epsilon$ brane. One can check that the two counterterms are mapped one into the other by the duality \eqref{eq:duality5} by expressing it in components. The regularised actions are then:
\begin{gather}
    \label{eq:reg_min_actions}
     S_{\sigma, \text{reg}} \supset  \int d^4 x \, A^{(0)}_{\mu} \pars{A_{(2)}^{\mu} - \frac{1}{4} \partial_\alpha F_{(0)}^{\alpha\mu} }, \quad
     S_{\rho, \text{reg}} \supset \frac{1}{4} \int d^4 x \, B^{(0)}_{\mu\nu} B_{(1)}^{\mu\nu}.
\end{gather}
Then, the currents can be identified by taking the $r\to\infty \; (u\to0)$ limit of \eqref{eq:membranes}, adapted to the presence of counterterms:
\begin{equation}
    \label{eq:min_currents}
    \langle j^\mu \rangle_\sigma = 2 A^\mu_{(2)} -  \frac{1}{2} \partial_\alpha F^{\alpha\mu}_{(0)}, \quad \langle \mathcal J^{\mu\nu} \rangle_r = B^{\mu\nu}_{(1)}.
\end{equation}
where I used subscripts in the expectation values to denote through which gravitational theory each of them is calculated. Comparing \eqref{eq:min_currents} with \eqref{eq:reg_min_actions} one can identify the corresponding external sources, and retrieve the holographic Kubo formulae. If the background state is assumed to be isotropic, $\sigma$ and $\rho$ are scalars that can be evaluated by considering charge dynamics -- i.e. bulk perturbations -- in a single spatial direction. Assuming that to be $x$, one finds:
\begin{equation}
    \label{eq:holo_kubo}
    \sigma(\omega) = \frac{2 \delta A^{(2)}}{i\omega \delta A^{(0)}} + \frac{i \omega}{2}, \quad
    \rho (\omega) = \frac{\delta B^{(1)}}{i \omega \delta B^{(0)} },
\end{equation}
where the FG coefficients appear in the asymptotic expansions of $\delta A_x$ and $\delta B_{yz}$.

It may appear odd that the conductivity $\sigma$ is made of two terms. This is a consequence of the FG expansion \eqref{eq:A_FG} having a logarithmic term, which is why the second term is absent in $d=3$ (or in any odd number of dimensions). However, while such a term is clearly irrelevant as far as DC ($\omega\to0$) transport is concerned, it has a dramatic impact at finite frequency (AC), and a clear phenomenological interpretation. Assuming the external source $A_{(0)}$ to obey Maxwell's equation in vacuum, the expectation value of the current in \eqref{eq:min_currents} reads:
\begin{equation}
    \langle j^\mu \rangle_\sigma = 2 A^\mu_{(2)} +  \frac{1}{2} j^{\mu}_\text{ext},
\end{equation}
where $j_\text{ext}$ is the electric current needed to support the chosen $A_{(0)}$ profile. This suggests that neglecting the extra term results in considering just part of the total current -- the one made of dynamical degrees of freedom. This difference is, of course, very relevant to the topic of reciprocity, as I will show later. As for the extra factor of 2 appearing in front of $j_\text{ext}$, a possible explanation is that the field lines it sources live in the half space $u>0$, thus making Coulomb interactions stronger by a factor of 2. This phenomenon has been noted -- albeit in one dimension less -- in QED\textsubscript{3,4}, as noted in \cite{Hsiao_Son}. A deeper analysis, which I leave to future work, would be needed to pinpoint the nature of this numerical prefactor precisely.

\paragraph{Alternative renormalisation schemes}

In the previous paragraph I anticipated that the minimal renormalisation procedure is not unique. Following \cite{GrozPoov_holoHF}, divergences could be in principle tamed by introducing the alternative counterterms:
\begin{equation}
    S_{\sigma, \text{alt-ct}} = - \frac{1}{4} \log \frac{\epsilon}{\mathcal C} \int_{u = \epsilon} d^4 x \, F^{(0)}_{\mu\nu} F_{(0)}^{\mu\nu}, \quad S_{\rho, \text{alt-ct}} = - \frac{1}{4} \log \frac{\epsilon}{\mathcal C} \int_{u = \epsilon} d^4 x \, \mathcal H_{\mu\nu} \mathcal H^{\mu\nu},
\end{equation}
where $\mathcal C$ is some arbitrary dimensionless number. Actions regularised in this way read:
\begin{equation}
   \begin{gathered}
        \label{eq:alt_regact}
        S_{\sigma, \text{alt-reg}} \supset  \int d^4 x \, A^{(0)}_{\mu} \spars{A_{(2)}^{\mu} - \frac{1}{2}\pars{\log{\mathcal C} + \frac{1}{2}} \partial_\alpha F_{(0)}^{\alpha\mu} }, \\[4pt]
        S_{\rho, \text{alt-reg}} \supset \frac{1}{4} \int d^4 x \, \pars{B^{(0)}_{\mu\nu} + B^{(1)}_{\mu\nu} \log{ \mathcal C}} B_{(1)}^{\mu\nu},
   \end{gathered} 
\end{equation}
and lead to alternative Kubo formulae:
\begin{equation}
    \label{eq:alt_kubo}
    \sigma(\omega) = \frac{2 \delta A^{(2)}}{i\omega \delta A^{(0)}} + i \omega \pars{\log {\mathcal C} + \frac{1}{2}}, \quad
    \rho(\omega) = \frac{\delta B^{(1)}}{i \omega \pars{\delta B^{(0)} + \delta B^{(1)} \log {\mathcal C}}}.
\end{equation}
I will refer to $\log \mathcal{C}$ as the \textit{renormalisation parameter}. Its introduction generalises the formulae \eqref{eq:holo_kubo}, to which \eqref{eq:alt_kubo} reduce upon choosing $\mathcal C = 1$. The freedom to pick any other value can be interpreted as the necessity to specify the energy scale under consideration, as proposed in \cite{GrozPoov_holoHF}. This is consistent with the fact that for any $\mathcal C \neq 1$, the regularised actions \eqref{eq:alt_regact} contain a double-trace operator, which breaks conformal invariance on the boundary. For a discussion of these kind of operators in holography, I refer to the discussion in \cite{Witten_multitrace}.

\paragraph{Anisotropy} Both \eqref{eq:holo_kubo} and \eqref{eq:alt_kubo} have been derived under the simplifying restriction to isotropic background states, which factorises charge dynamics into distinct directions. As I shall show later, however, the reciprocity proof applies to a larger class of states. From the viewpoint of linear response theory, anisotropy allows sources to generate non-zero responses in orthogonal directions as well. In the gravitational dual picture, this corresponds to a decrease in the number of independent channels in the perturbative equations. This makes the tensor structures of $\sigma$ and $\rho$ more complicated, as the holographic Kubo formulae become:
\begin{equation}
    \sigma(\omega)_{ij} = \frac{2 \delta A_i^{(2)}}{i\omega \delta A_j^{(0)}} + i \omega \pars{\log {\mathcal C} + \frac{1}{2}} \delta_{ij}, \quad
    \rho(\omega)_{ij} =  \frac{\varepsilon_{ikl} \delta B^{(1)}_{kl}}{i \omega \varepsilon_{jmn} \pars{\delta B^{(0)}_{mn} + \delta B^{(1)}_{mn} \log {\mathcal C}}},
\end{equation}
with $\varepsilon_{xyz} =1$ being the totally antisymmetric symbol in $3d$ space. The number of independent coefficients in the formulae is determined by residual symmetry. The introduction of a magnetic field in a neutral background, for instance, yields to two independent resistivities, as was shown explictily in \cite{GrozPoov_holoHF}.

\section{An analytic proof of reciprocity}
\label{sec:analytic_proof}

In this section I provide an analytic proof of the reciprocity relation, and discuss its validity. As spatial momentum is equal to 0 in all Kubo formulae considered so far, I will limit my discussion to the case of homogeneous, alternating perturbations on the boundary. I work in FG coordinates, so that the metric tensor is given by \eqref{FGmetric_x}.

The most general bulk perturbations (on both sides of \eqref{eq:duality5}) are:
\begin{gather}
    \delta A = \pars{ \delta A_x (u) dx + \delta A_y (u) dy + \delta A_z (u) dz }  e^{-i \omega t} , \\[4pt]
    \delta B = \pars{ \delta B_{xy} (u) dx\wedge dy + \delta B_{yz} (u) dy \wedge dz + \delta B_{zx} (u) dz\wedge dx } e^{-i \omega t} ,
\end{gather}
which in terms of tensor strengths are:
\begin{gather}
    \label{eq:deltaF}
    \delta F = \pars{ \delta A_x' (u) du \wedge dx -  i \omega \delta A_x (u) dt \wedge dx }e^{-i \omega t} + (x\to y) + (x\to z) ,\\[4pt]
    \begin{split}
    \label{eq:deltaH}
    \delta H = \pars{ \delta B_{xy}' (u) du \wedge dx \wedge dy -  i \omega \delta B_{xy} (u) dt \wedge dx \wedge dy} e^{-i \omega t} + \\ + (xy\to yz) + (xy\to zx).
    \end{split}
\end{gather}
The dualisation \eqref{eq:duality5} acts on the 2-forms in \eqref{eq:deltaF} as:
\begin{equation}
    \begin{gathered}
        *_5 (du\wedge dx) = \frac{\sqrt{-g}}{3!} g^{uu} g^{x\mu} \varepsilon_{u \mu \nu \sigma \rho} dx^\nu \wedge dx^\sigma \wedge dx^\rho \simeq  \frac{1}{u} dt \wedge dy \wedge dz + O(u), \\[4pt]
        *_5 (dt\wedge dx) = \frac{\sqrt{-g}}{3!} g^{ta} g^{xb} \varepsilon_{abcde} dx^a \wedge dx^b \wedge dx^c \simeq  \frac{1}{u} du \wedge dy \wedge dz + O(u),
    \end{gathered} 
\end{equation}
where $\varepsilon_{txyzu}=-1$ is the antisymmetric symbol in $5d$, and the second (approximate) equality holds as long as the leading coefficient in the Fefferman-Graham expansion for $\gamma_{\mu\nu}$ is $\eta_{\mu\nu}$ -- i.e., if the spacetime under consideration is asymptotically AdS. Analogous expressions can be found for $du\wedge dy$, $dt\wedge dy$, $du\wedge dz$ and $dt\wedge dz$. Through these and \eqref{eq:deltaF}, one finds an alternative expression for $\delta H$, which reads, close to the AdS boundary:
\begin{equation}
    \begin{aligned}
    \delta H = \bigg(\frac{1}{u} \delta A_x' (u) dt \wedge dy \wedge dz - \frac{i \omega}{u} \delta A_x(u) du\wedge dy\wedge dz + & \\ + (xyz \to yzx) + (xyz \to zxy)\bigg) & e^{-i \omega t} + O(u).
    \end{aligned}
\end{equation}
Equating the result with \eqref{eq:deltaH} leads then to:
\begin{equation}
    \begin{gathered}
        \label{eq:mhb_constraints}
        - i \omega \delta B_{yz}(u) = \frac{1}{u} \delta A_x'(u) + O(u), \quad 
        \delta B_{yz}'(u) = - \frac{i \omega}{u} \delta A_x(u) + O(u),
    \end{gathered}
\end{equation}
and four additional relations obtained by considering cyclic permutations of $x,y$ and $z$. Plugging the FG expansions \eqref{eq:A_FG} and \eqref{eq:B_FG} into \eqref{eq:mhb_constraints} leads to:
\begin{equation}
    \begin{gathered}
        -i \omega \pars { \delta B^{(0)}_{yz} + \delta B^{(1)}_{yz} \log{u} } = 2 \delta A_x^{(2)} - \omega^2 \pars{\log{u} + \frac{1}{2}} \delta A_x^{(0)} + O(u \log{u}), \\
        \delta B_{yz}^{(1)} = - i \omega \delta A_x^{(0)} + O(u \log{u}),
    \end{gathered}
\end{equation}
and analogous expressions for cyclic permutations of $x,y$ and $z$. This in turn leads to the identifications (at $u=0$):
\begin{equation}
    \label{eq:bdy_rels}
    \begin{gathered}
        \delta B_{yz}^{(0)} = \frac{2i}{\omega} \delta A_x^{(2)} - \frac{i \omega}{2} \delta A_{x}^{(0)}, \quad 
        \delta B_{yz}^{(1)} = - i \omega \delta A_x^{(0)}, \\[4pt]
        \delta B_{zx}^{(0)} = \frac{2i}{\omega} \delta A_y^{(2)} - \frac{i \omega}{2} \delta A_{y}^{(0)}, \quad 
        \delta B_{zx}^{(1)} = - i \omega \delta A_y^{(0)}, \\[4pt]
        \delta B_{xy}^{(0)} = \frac{2i}{\omega} \delta A_z^{(2)} - \frac{i \omega}{2} \delta A_{z}^{(0)}, \quad 
        \delta B_{xy}^{(1)} = - i \omega \delta A_z^{(0)}.
    \end{gathered}
\end{equation}
These relations do not follow from a specific form of the action or background: they are just a manifestation of the bulk duality \eqref{eq:duality5}. The only requirement is that the metric \eqref{FGmetric_x} asymptotes to AdS. Substitution into \eqref{eq:alt_kubo} easily leads to the reciprocity relation for arbitrary values of the renormalisation parameter $\log \mathcal C$ in the scalar case; the anisotropic case requires some more work -- as the functional derivatives need to be re-expressed as suitable ratios of FG coefficients -- but leads to the same result. The conclusion is that:
\begin{equation}
    \label{eq:ohm_nice}
    \sigma_{ij}(\omega) \rho_{jk}(\omega) = \delta_{ik}.
\end{equation}
I will postpone a discussion on the regime of validity of this formula to Section \ref{sec:disc}. Before doing so, however, I confirm its validity with a numerical study of the backgrounds described in Section \ref{sec:bcg}.

\section{Numerical transport coefficients}
\label{sec:tc}

This section is dedicated to the numerical calculation of the AC transport coefficients for the backgrounds described in Section \ref{sec:bcg}. As these are isotropic, one can consider charge dynamics along a single direction (which I will pick to be the $x$ axis); the appropriate definitions for $\sigma$ and $\rho$ are given in \eqref{eq:alt_kubo}.

As for the conductivity $\sigma$, one starts by considering the following perturbations:
\begin{equation}
    \label{eq:perts_sigma}
    \delta \psi^x = \delta \psi(r) e^{-i \omega t}, \quad \delta A = \delta a(r) e^{-i \omega t} dx, \quad \delta (ds^2) = 2 r^2 \delta h(r) e^{-i \omega t} dt dx.
\end{equation}
Plugging these expressions into \eqref{eq:aw_covariant_eoms} gives a set of radial ODEs that have been extensively studied the past: in \cite{aw_axions}, they have been used to derive an analytic formula for the DC ($\omega \to 0$) conductivity $\sigma_{\scalebox{.6}{\text{DC}}}$, whereas the numerical result for the frequency-dependent AC conductivity can be found in \cite{Kim_num}. I will not discuss these equations, and quote the result when discussing the resistivity. More detail on the physics than can be extracted from those simple equations can be found in \cite{axion_review}. \medskip

Similarly, in order to calculate the holographic resistivity $\rho$, it is convenient to consider charge dynamics along the $x$ direction. The relevant perturbations are, in the radial gauge:
\begin{equation}
    \label{eq:perts_r}
    \delta \psi^x = \delta \psi(r) e^{-i \omega t}, \quad \delta B = \delta b(r) e^{-i \omega t} dy \wedge dz, \quad \delta (ds^2) = 2 r^2 \delta h(r) e^{-i \omega t} dt dx.
\end{equation}
Plugging these into \eqref{eq:hf_covariant_eoms} one finds:
\begin{gather}
    \label{eq:r_eom1}
    \delta \psi '' + \pars{\frac{f'}{f} + \frac{3}{r}} \delta \psi' + \frac{\omega^2}{f^2} \delta \psi - \frac{i \alpha \omega}{f^2} \delta h = 0, \\[4pt]
    \label{eq:r_eom2}
    \delta b '' + \pars{\frac{f'}{f} - \frac{1}{r}} \delta b' + \frac{\omega^2}{f^2} \delta b - \frac{i n \omega}{f^2} \delta h = 0, \\[4pt]
    \label{eq:r_eom3}
    \alpha \delta \psi ' + \frac{n}{r^4} \delta b' - \frac{i r^2 \omega}{f} \delta h' = 0,
\end{gather}
where primes denote a radial derivative, plus a fourth equation containing $\delta h''$, which is implied by the ones above. As expected, setting the charge density $n$ to 0 decouples the electromagnetic perturbation $\delta b$ from the rest. Equations \eqref{eq:r_eom1}-\eqref{eq:r_eom3} are superficially similar to the ones reported in \cite{aw_axions}, but their structure -- which encodes a different asymptotic expansion -- is different enough to make it impossible to repeat their argument for an analytic DC formula.

The task of numerically integrating \eqref{eq:r_eom1}-\eqref{eq:r_eom3} is made simpler by picking a radial coordinate with a compact support. An obvious candidate is $z = r_h / r$.\footnote{This is not to be confused with the spatial boundary coordinate $z$. Context allows to distinguish the two of them without ambiguities.} The AdS boundary is at $z=0$ and the horizon at $z=1$. Notice that $z\neq u$, the latter being the FG radial coordinate, as it can seen by substituting $r$ for $z$ in \eqref{eq:SStoFG}. In this new coordinate system the equations of motion become:
\begin{gather}
    \label{eq:z_eom1}
    \delta \psi '' + \pars{\frac{f'}{f} - \frac{1}{z}} \delta \psi' + \frac{r_h^2 \omega^2}{z^4 f^2} \delta \psi - \frac{i r_h^2 \alpha \omega}{z^4 f^2} \delta h = 0, \\[4pt]
    \label{eq:z_eom2}
    \delta b '' + \pars{\frac{f'}{f} + \frac{3}{z}} \delta b' + \frac{r_h^2 \omega^2}{z^4 f^2} \delta b - \frac{i r_h^2 n \omega}{z^4 f^2} \delta h = 0, \\[4pt]
    \label{eq:z_eom3}
    \alpha \delta \psi ' + \frac{z^4 n}{r_h^4} \delta b' - \frac{i r_h^2 \omega}{z^2 f} \delta h' = 0,
\end{gather}
with primes now denoting $z$ derivatives. Boundary conditions can be imposed by studying the IR and UV asymptotics of the equations. In the former case, solutions are required to be radially infalling near the horizon:
\begin{equation}
    \begin{gathered}
        \label{eq:IR_exp}
        \delta \psi \sim (1-z)^{- \frac{i \omega}{4 \pi T}} \sum_{n} \delta \psi_{\scalebox{.7}{\text{IR}}}^{(n)}(1-z)^n, \quad \delta b \sim (1-z)^{- \frac{i \omega}{4 \pi T}} \sum_{n} \delta b_{\scalebox{.7}{\text{IR}}}^{(n)}(1-z)^n, \\[4pt]
        \delta h \sim (1-z)^{- \frac{i \omega}{4 \pi T} + 1} \sum_{n} \delta h_{\scalebox{.7}{\text{IR}}}^{(n)}(1-z)^n.
    \end{gathered}
\end{equation}
Close to the UV boundary, the asymptotic expansion must take into account of possible logarithmic terms. A simple ansatz is then:
\begin{equation}
    \begin{gathered}
        \label{eq:UV_exp}
        \delta \psi \sim \sum_{n} z^n \pars{\delta \psi_{\scalebox{.7}{\text{UV}}}^{(n)} + \delta \tilde \psi_{\scalebox{.7}{\text{UV}}}^{(n)} \, \log z  } , \quad
        \delta b \sim \sum_{n} z^n \pars{\delta b_{\scalebox{.7}{\text{UV}}}^{(n)} + \delta \tilde b_{\scalebox{.7}{\text{UV}}}^{(n)} \, \log z  }, \\[4pt]
        \delta h \sim \sum_{n} z^n \pars{\delta h_{\scalebox{.7}{\text{UV}}}^{(n)} + \delta \tilde h_{\scalebox{.7}{\text{UV}}}^{(n)} \, \log z  }.
    \end{gathered}
\end{equation}
By inserting the expansions \eqref{eq:IR_exp} and \eqref{eq:UV_exp} into the equations of motion \eqref{eq:z_eom1}-\eqref{eq:z_eom3}, one finds that all coefficients are either 0 or recursively determined from:
\begin{equation}
    \label{eq:free_coeffs}
    \delta \psi_{\scalebox{.7}{\text{IR}}}^{(0)}, \, \delta b_{\scalebox{.7}{\text{IR}}}^{(0)}, \, \delta \psi_{\scalebox{.7}{\text{UV}}}^{(0)}, \, \delta b_{\scalebox{.7}{\text{UV}}}^{(0)}, \, \delta h_{\scalebox{.7}{\text{UV}}}^{(0)}, \, \delta \psi_{\scalebox{.7}{\text{UV}}}^{(4)}, \,  \delta \tilde b_{\scalebox{.7}{\text{UV}}}^{(0)}.
\end{equation}
There are thus 2 coefficients more than expected, as \eqref{eq:z_eom1}-\eqref{eq:z_eom3} form a system of 2 second order and 1 first order ODEs, the solutions of which are uniquely determined by 5 boundary conditions. The redundancy can be fixed by noticing that the equations of motion are invariant under the two translational symmetries:
\begin{equation}
    \label{eq:transl}
    \delta\psi \to \delta\psi + c_1, \quad \delta b \to \delta b + c_2, \quad \delta h \to \delta h - i \omega \pars{ \frac{c_1}{\alpha} + \frac{c_2}{n} },
\end{equation}
where $c_1$ and $c_2$ are arbitrary constants. These symmetries can be understood in terms of a residual gauge invariance that survives the radial gauge choice in \eqref{eq:perts_r}; to see this, it is sufficient to notice that it is the action of the diffeomorphism generated by: $\xi^a = \delta^a_r \xi(r) e^{-i \omega t}$ that allows to set $\delta g_{rx}=0$ in the first place. \eqref{eq:transl} are the transformations generated by $\xi(r)=\text{const}$. An analogous phenomenon, which does not involve all three fields, has been extensively reported in calculations of the conductivity, starting from \cite{Qlattices}.

Because the equations of motion \eqref{eq:z_eom1}-\eqref{eq:z_eom3} are linear in the perturbations, rescaling the fields by the same constant factor leaves them invariant. All these considerations, then, inform a strategy for the numerical calculation of the resistivity $\rho$:
\begin{itemize}
    \item select a background state identified by $(\alpha/T,n/T^3)$;
    \item start integrating \eqref{eq:z_eom1}--\eqref{eq:z_eom3} numerically from $z=1$ towards $z=0$, fixing one of the two free IR coefficients to 1 and leaving the other free;
    \item fix the latter by requiring the combination $\omega \delta \psi_{\scalebox{.7}{\text{UV}}}^{(0)} - i \alpha \delta h_{\scalebox{.7}{\text{UV}}}^{(0)}$ to vanish, and set $\delta h_{\scalebox{.7}{\text{UV}}}^{(0)}=0$ via \eqref{eq:transl}. This ensures that only the electromagnetic source is turned on;
    \item read the resistivity off at the boundary $z=0$ using \eqref{eq:alt_kubo}.
\end{itemize}

\begin{figure}
    \centering
    \includegraphics[width = \linewidth]{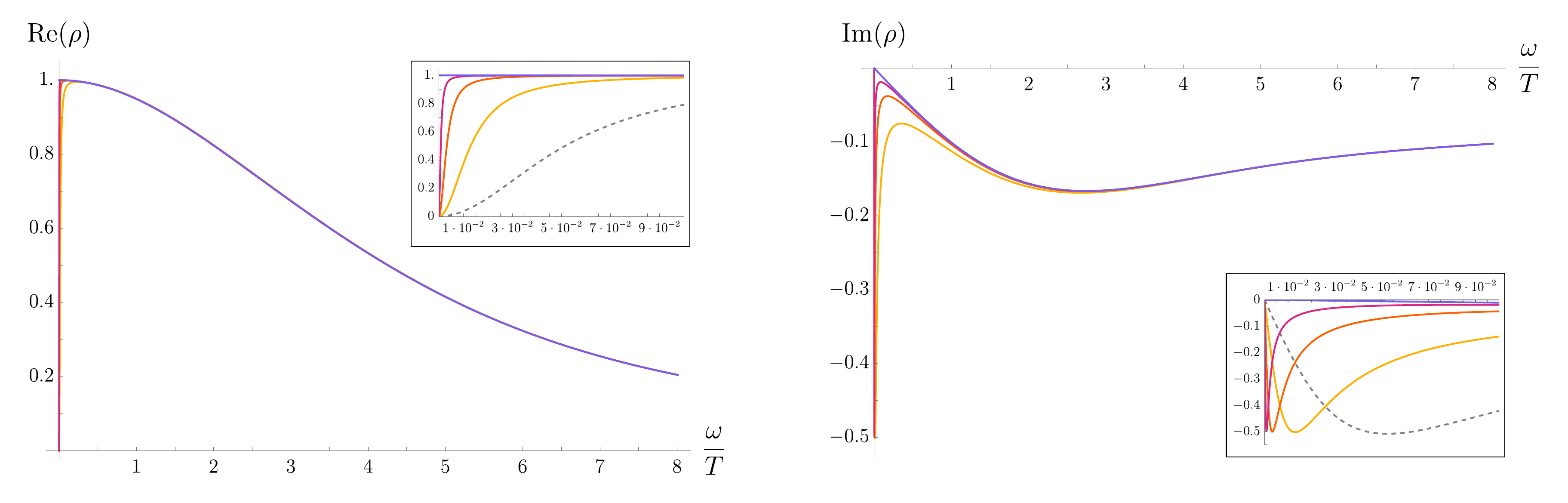}
    \caption{AC resistivity in absence of momentum dissipation for different values of $n/T^3$: 0 (violet), 1 (magenta), 2 (orange) and 4 (yellow). In the insets -- which depict the small frequency behavior -- the additional value 8 (grey, dashed) has been plotted.}
    \label{fig:rho-var}
\end{figure}

\begin{figure}
    \centering
    \includegraphics[width = \linewidth]{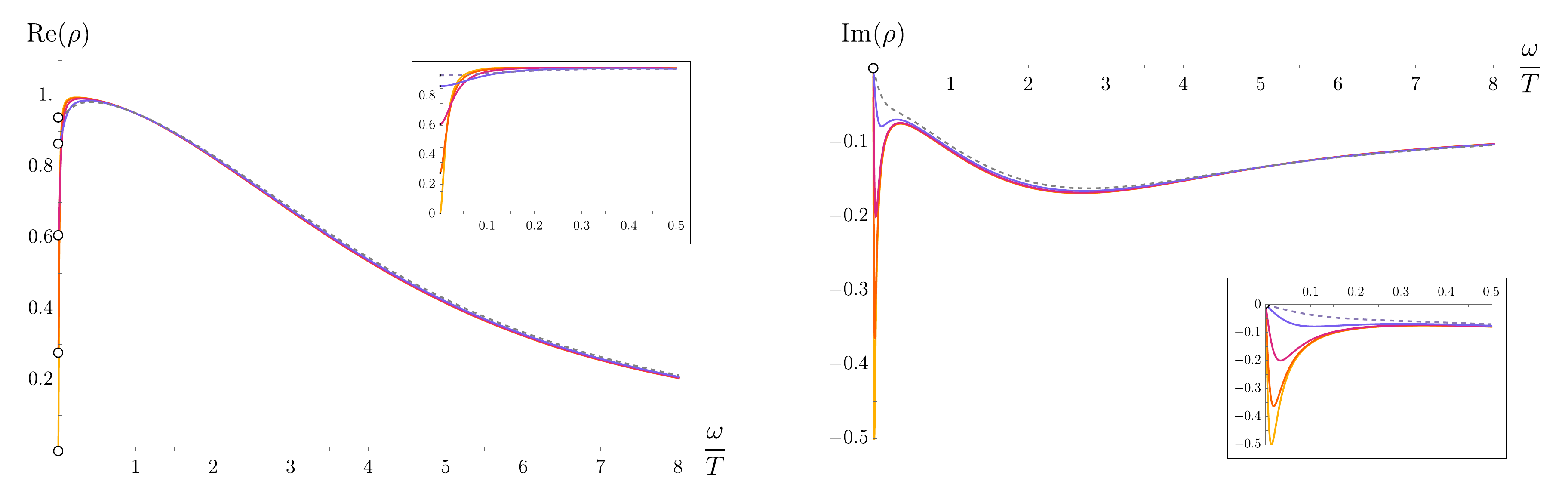}
    \caption{AC resistivity at $n/T^3=4$ for different values of $\alpha/T$: 0 (yellow), 0.25 (orange), 0.5 (magenta), 1 (violet) and 1.5 (grey, dashed). The circles are the inverses of the appropriate analytic DC prediciton of \cite{aw_axions}. Insets depict the small frequency behavior of the curves.}
    \label{fig:alpha-var}
\end{figure}

\begin{figure}
    \centering
    \includegraphics[width = \linewidth]{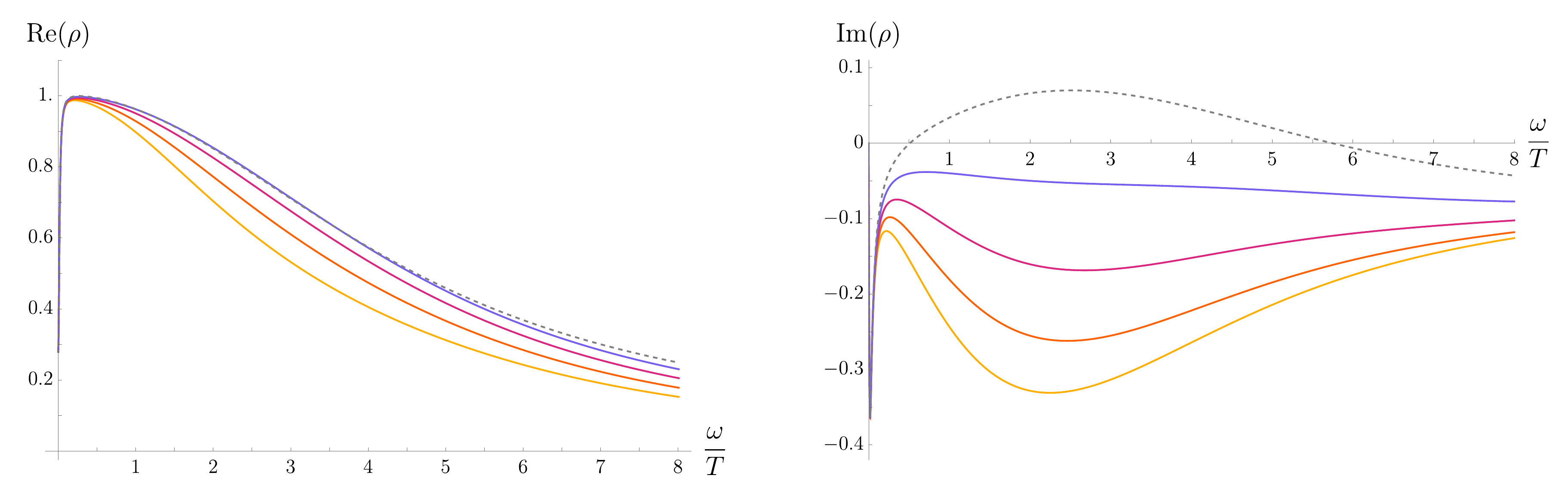}
    \caption{AC resistivity at $n/T^3 = 4, \alpha/T = 0.25$ for different value of the renormalisation parameter $\log \mathcal C$: 0.5 (yellow), 0.25 (orange), 0 (magenta), -0.25 (violet), -0.5 (grey, dashed). The last value corresponds to calculate $\sigma$ without external contribution -- see \eqref{eq:holo_kubo} and the discussion that follows.}
    \label{fig:lp-var}
\end{figure}

The outcome of numerical calculations is summarised in Figures \ref{fig:rho-var}, \ref{fig:alpha-var} and \ref{fig:lp-var}; unless otherwise stated, the calculations have been performed with $\log\mathcal C = 0$. In all of the cases under consideration, inverse conductivities and resistivities are indistinguishable to the precision of the integration algorithms used. This constitutes direct evidence in support of the analytic argument presented above.

On top of that, a number of other features comply with predictions valid for the holographic conductivity. Figure \ref{fig:rho-var} shows that in absence of momentum dissipation ($\alpha = 0$) $\rho_{DC} = 0$ whenever $n\neq 0$. This is consistent with the fact that in such states the imaginary part of the holographic conductivity $\sigma$ diverges for small frequencies, as argued in \cite{Hartnoll-lectures}. Unlike it, in addition, the real part of $\rho$ does go smoothly to its expected DC value, which has to be 0 on the grounds of translational symmetry.

Such symmetry can be broken by setting a finite value to $\alpha/T$; for those states, the DC conductivity takes the analytic value calculated in \cite{aw_axions}. The numerical AC resistivities approach smoothly, in the DC limit, its inverse. This is depicted in Figure \ref{fig:alpha-var}, which suggests that the stronger the momentum dissipation, the more resistive the phase under consideration. This is intuitive and reminiscent of the Joule effect. Another notable effect is the disappearance of the one of the two minima of the imaginary part as momentum dissipation increases.

Lastly, I consider the effect of a non-zero renormalisation parameter $\log\mathcal C$ on the  transport coefficients. AC resistivities at different energy scales are reported in Figure \ref{fig:lp-var}. An interesting feature is the fact that, while positive values of $\log \mathcal C$ make the imaginary part of $\rho$ more negative -- and the system less resistive --, there appears to be a critical negative value in the interval $(-0.5,-0.25)$ below which $\text{Im}(\rho)>0$ for some finite range of frequencies. It is also worth noticing that the different behavior of the magenta and dashed, grey curves in Figure \ref{fig:lp-var} is direct evidence of the fact that the inverse conductivity \eqref{eq:holo_kubo} without the second term does not invert to the correspondent resistivity.

\section{Discussion}
\label{sec:disc}

The main result of this work is the formula \eqref{eq:ohm_nice}, which I derived analytically and confirmed numerically in a number of simple settings. Another important milestone is the dualisation criterion enunciated towards the end of Section \ref{sec:bcg}. These inform two possible future developments: the search for deviations from \eqref{eq:ohm_nice} in holography, and the existence of loopholes to the contradiction in \eqref{eq:contradiction}.

Regarding the former, the result does not seem to heavily depend on the specific form of the actions \eqref{eq:aw_action} and \eqref{eq:hf_action}, but on the dualisation relation relating them. This is clarified by the calculation I presented, in which the equations and boundary conditions that lead to the specific profile of either $\delta A$ or $\delta B$ play no role; the actions enter the discussion only through the Kubo formulae, which are fixed by the quadratic nature of the electromagnetic part of the action. The expectation is then that any two actions, quadratic in $F$, $H$ and related by \eqref{eq:duality5} -- i.e., avoiding the contradiction \eqref{eq:contradiction} -- should produce the same result. There are nonetheless a few extra conditions that need to be met in order to obtain the result:
\begin{itemize}
    \item the metric must be asymptotically AdS close to the UV boundary;
    \item the boundary currents must follow from \eqref{eq:membranes};
    \item the bulk must be entirely described by classical fields.
\end{itemize}
The last observation suggests that stringy corrections may, at least in principle, produce a deviation from \eqref{eq:ohm_nice}. From the boundary viewpoint, this would amount to considering finite-$N$ corrections, which have been shown in \cite{toapp} to be very relevant to the validity of the statement. Concretely, this could be achieved by considering curvature-square corrections to Einstein gravity, for instance by adding to the action a Gauss-Bonnet-type term, as in \cite{Grozdanov_GB,Folkestad_2019}.

The conditions listed above are nevertheless met by a large class of holographic models. One such example are the the magnetic brane solutions studied in \cite{DHK_nMB,DHK_cMB}, which describe boundary states with uniform background magnetic fields. These are of particular interest as they offer a simple tool to study properties of matter in strong magnetic fields. This is a notoriously hard task if undertaken with conventional methods -- see \cite{Vardhan:2022wxz,Vardhan:2024qdi} for a new approach with novel EFT techniques --, but of the utmost importance to shed some light into the phenomenology of nuclear matter in neutron stars, and possibly -- by considering different horizon topologies -- to give insights into Tokamak physics. Specifically, one could investigate whether gravitational instabilities in the bulk translate into plasma instabilities on the boundary \cite{instab_plasma1,mikhailovskii2017instabilities}.

On the other hand, exceptions to the contradiction \eqref{eq:contradiction} at the heart of the proposed criterion for dualisation could come from a modification of the map \eqref{eq:duality5}, and allowing violations of the Bianchi identities $dX=0$. In $D=3+1$ bulk dimensions, this can be understood in terms of the introduction of magnetic monopoles in the dual theory; in $D=4+1$, no such clear interpretation is readily available. Regardless, advances on this front could eventually lead to a procedure to dualise holographic models with $U(1)$-charged matter.

Another possible direction could be to find holographic models allowing the investigation of Ohm's reciprocity in the strict sense -- i.e. allowing for the simultaneous calculation of both $\sigma$ and $\rho$. A possibility could be to merge \eqref{eq:aw_action} and \eqref{eq:hf_action} into a single action. In this case, the duality \eqref{eq:duality5} could perhaps become a self-duality, in analogy to the lower-dimensional case studied in \cite{Mtheory}. A substantial difference between the two cases is nevertheless that, in $d=2+1$ boundary dimensions, the kinetic term of the photon is irrelevant and removed by flowing to the IR -- see also \cite{Giombi_confCFT} --, so that the dualisation does not introduce dynamical electromagnetism, and the resulting self-dual model is considerably simpler.

\acknowledgments

I would like to thank my doctoral supervisor Sa\v{s}o Grozdanov and my fellow student Mile Vrbica for interesting conversations we had while I was working on this project, and comments on an early version of this manuscript. I am also grateful to Matteo Baggioli who stoked my interest in the topic. My doctoral studies are supported by an Edinburgh Doctoral College Scholarship (ECDS) from the University of Edinburgh.

% Bibliography

%% [A] Recommended: using JHEP.bst file
\bibliographystyle{JHEP}
\bibliography{biblio.bib}

\providecommand{\href}[2]{#2}\begingroup\raggedright\begin{thebibliography}{10}

\bibitem{kittel}
C.~Kittel, \emph{Introduction to Solid State Physics}, Wiley (2004).

\bibitem{Hernandez_2017}
J.~Hernandez and P.~Kovtun, \emph{Relativistic magnetohydrodynamics},
  \href{https://doi.org/10.1007/jhep05(2017)001}{\emph{Journal of High Energy
  Physics} {\bfseries 2017} (2017) }.

\bibitem{Kubo:original}
R.~Kubo, \emph{{Statistical mechanical theory of irreversible processes. 1.
  General theory and simple applications in magnetic and conduction problems}},
  \href{https://doi.org/10.1143/JPSJ.12.570}{\emph{J. Phys. Soc. Jap.}
  {\bfseries 12} (1957) 570}.

\bibitem{AMYpaper}
P.~Arnold, G.D.~Moore and L.G.~Yaffe, \emph{Transport coefficients in high
  temperature gauge theories (i): leading-log results},
  \href{https://doi.org/10.1088/1126-6708/2000/11/001}{\emph{Journal of High
  Energy Physics} {\bfseries 2000} (2000) 001}.

\bibitem{Kovtun:2012rj}
P.~Kovtun, \emph{{Lectures on hydrodynamic fluctuations in relativistic
  theories}}, \href{https://doi.org/10.1088/1751-8113/45/47/473001}{\emph{J.
  Phys. A} {\bfseries 45} (2012) 473001}
  [\href{https://arxiv.org/abs/1205.5040}{{\ttfamily 1205.5040}}].

\bibitem{KapustaGale}
J.I.~Kapusta and C.~Gale, \emph{{Finite-temperature field theory: Principles
  and applications}}, Cambridge Monographs on Mathematical Physics, Cambridge
  University Press (2011),
  \href{https://doi.org/10.1017/CBO9780511535130}{10.1017/CBO9780511535130}.

\bibitem{GHI_2formmhd}
S.~Grozdanov, D.M.~Hofman and N.~Iqbal, \emph{{Generalized global symmetries
  and dissipative magnetohydrodynamics}},
  \href{https://doi.org/10.1103/PhysRevD.95.096003}{\emph{Phys. Rev. D}
  {\bfseries 95} (2017) 096003}
  [\href{https://arxiv.org/abs/1610.07392}{{\ttfamily 1610.07392}}].

\bibitem{toapp}
G.~Frangi and S.~Grozdanov, \emph{{Quantum origin of Ohm's reciprocity relation
  and its violation: conductivity as inverse resistivity}},
  \href{https://arxiv.org/abs/2406.16123}{{\ttfamily 2406.16123}}.

\bibitem{Maldacena:1997re}
J.M.~Maldacena, \emph{{The Large N limit of superconformal field theories and
  supergravity}}, \href{https://doi.org/10.4310/ATMP.1998.v2.n2.a1}{\emph{Adv.
  Theor. Math. Phys.} {\bfseries 2} (1998) 231}
  [\href{https://arxiv.org/abs/hep-th/9711200}{{\ttfamily hep-th/9711200}}].

\bibitem{Jeon:long}
S.~Jeon, \emph{{Hydrodynamic transport coefficients in relativistic scalar
  field theory}}, \href{https://doi.org/10.1103/PhysRevD.52.3591}{\emph{Phys.
  Rev. D} {\bfseries 52} (1995) 3591}
  [\href{https://arxiv.org/abs/hep-ph/9409250}{{\ttfamily hep-ph/9409250}}].

\bibitem{Jeon:Yaffe}
S.~Jeon and L.G.~Yaffe, \emph{{From quantum field theory to hydrodynamics:
  Transport coefficients and effective kinetic theory}},
  \href{https://doi.org/10.1103/PhysRevD.53.5799}{\emph{Phys. Rev. D}
  {\bfseries 53} (1996) 5799}
  [\href{https://arxiv.org/abs/hep-ph/9512263}{{\ttfamily hep-ph/9512263}}].

\bibitem{Damle_1997}
K.~Damle and S.~Sachdev, \emph{Nonzero-temperature transport near quantum
  critical points},
  \href{https://doi.org/10.1103/physrevb.56.8714}{\emph{Physical Review B}
  {\bfseries 56} (1997) 8714–8733}.

\bibitem{Sachdev_1998}
S.~Sachdev, \emph{Nonzero-temperature transport near fractional quantum hall
  critical points},
  \href{https://doi.org/10.1103/physrevb.57.7157}{\emph{Physical Review B}
  {\bfseries 57} (1998) 7157–7173}.

\bibitem{SS_minkPrescr}
D.T.~Son and A.O.~Starinets, \emph{{Minkowski-space correlators in AdS/CFT
  correspondence: recipe and applications}},
  \href{https://doi.org/10.1088/1126-6708/2002/09/042}{\emph{Journal of High
  Energy Physics} {\bfseries 2002} (2002) 042–042}.

\bibitem{Hall_HK}
S.A.~Hartnoll and P.K.~Kovtun, \emph{Hall conductivity from dyonic black
  holes}, \href{https://doi.org/10.1103/physrevd.76.066001}{\emph{Physical
  Review D} {\bfseries 76} (2007) }.

\bibitem{Hartnoll-lectures}
S.A.~Hartnoll, \emph{{Lectures on holographic methods for condensed matter
  physics}}, \href{https://doi.org/10.1088/0264-9381/26/22/224002}{\emph{Class.
  Quant. Grav.} {\bfseries 26} (2009) 224002}
  [\href{https://arxiv.org/abs/0903.3246}{{\ttfamily 0903.3246}}].

\bibitem{aw_axions}
T.~Andrade and B.~Withers, \emph{{A simple holographic model of momentum
  relaxation}}, \href{https://doi.org/10.1007/jhep05(2014)101}{\emph{Journal of
  High Energy Physics} {\bfseries 2014} (2014) }.

\bibitem{DHK_nMB}
E.~D’Hoker and P.~Kraus, \emph{{Magnetic brane solutions in AdS}},
  \href{https://doi.org/10.1088/1126-6708/2009/10/088}{\emph{Journal of High
  Energy Physics} {\bfseries 2009} (2009) 088–088}.

\bibitem{hartnoll2018holographic}
S.A.~Hartnoll, A.~Lucas and S.~Sachdev, \emph{Holographic quantum matter}, MIT
  Press (2018), [\href{https://arxiv.org/abs/1612.07324}{{\ttfamily
  1612.07324}}].

\bibitem{axion_review}
M.~Baggioli, K.-Y.~Kim, L.~Li and W.-J.~Li, \emph{Holographic axion model: A
  simple gravitational tool for quantum matter},
  \href{https://doi.org/10.1007/s11433-021-1681-8}{\emph{Sci. China Phys. Mech.
  Astr.} {\bfseries 64} (2021) }.

\bibitem{GrozPoov_holoHF}
S.~Grozdanov and N.~Poovuttikul, \emph{{Generalised global symmetries in
  holography: magnetohydrodynamic waves in a strongly interacting plasma}},
  \href{https://doi.org/10.1007/JHEP04(2019)141}{\emph{JHEP} {\bfseries 04}
  (2019) 141} [\href{https://arxiv.org/abs/1707.04182}{{\ttfamily
  1707.04182}}].

\bibitem{HofIqb_holoHF}
D.M.~Hofman and N.~Iqbal, \emph{{Generalized global symmetries and
  holography}},
  \href{https://doi.org/10.21468/SciPostPhys.4.1.005}{\emph{SciPost Phys.}
  {\bfseries 4} (2018) 005} [\href{https://arxiv.org/abs/1707.08577}{{\ttfamily
  1707.08577}}].

\bibitem{dynboundary}
S.~Grozdanov, A.~Lucas and N.~Poovuttikul, \emph{Holography and hydrodynamics
  with weakly broken symmetries},
  \href{https://doi.org/10.1103/physrevd.99.086012}{\emph{Physical Review D}
  {\bfseries 99} (2019) }.

\bibitem{DeWolfeHiggin}
O.~DeWolfe and K.~Higginbotham, \emph{{Generalized symmetries and 2-groups via
  electromagnetic duality in $AdS/CFT$}},
  \href{https://doi.org/10.1103/PhysRevD.103.026011}{\emph{Phys. Rev. D}
  {\bfseries 103} (2021) 026011}
  [\href{https://arxiv.org/abs/2010.06594}{{\ttfamily 2010.06594}}].

\bibitem{IqbalLiu_MB}
N.~Iqbal and H.~Liu, \emph{{Universality of the hydrodynamic limit in AdS/CFT
  and the membrane paradigm}},
  \href{https://doi.org/10.1103/PhysRevD.79.025023}{\emph{Phys. Rev. D}
  {\bfseries 79} (2009) 025023}
  [\href{https://arxiv.org/abs/0809.3808}{{\ttfamily 0809.3808}}].

\bibitem{Kovtun_holohydro}
P.~Kovtun, D.T.~Son and A.O.~Starinets, \emph{Holography and hydrodynamics:
  diffusion on stretched horizons},
  \href{https://doi.org/10.1088/1126-6708/2003/10/064}{\emph{Journal of High
  Energy Physics} {\bfseries 2003} (2003) 064–064}.

\bibitem{Hartnoll:hsc1}
S.A.~Hartnoll, C.P.~Herzog and G.T.~Horowitz, \emph{{Building a Holographic
  Superconductor}},
  \href{https://doi.org/10.1103/PhysRevLett.101.031601}{\emph{Phys. Rev. Lett.}
  {\bfseries 101} (2008) 031601}
  [\href{https://arxiv.org/abs/0803.3295}{{\ttfamily 0803.3295}}].

\bibitem{Hartnoll:hsc2}
S.A.~Hartnoll, C.P.~Herzog and G.T.~Horowitz, \emph{{Holographic
  Superconductors}},
  \href{https://doi.org/10.1088/1126-6708/2008/12/015}{\emph{JHEP} {\bfseries
  12} (2008) 015} [\href{https://arxiv.org/abs/0810.1563}{{\ttfamily
  0810.1563}}].

\bibitem{witten1998anti}
E.~Witten, \emph{{Anti-de Sitter space and holography}},
  \href{https://doi.org/10.4310/ATMP.1998.v2.n2.a2}{\emph{Adv. Theor. Math.
  Phys.} {\bfseries 2} (1998) 253}
  [\href{https://arxiv.org/abs/hep-th/9802150}{{\ttfamily hep-th/9802150}}].

\bibitem{Chamblin_1999}
A.~Chamblin, R.~Emparan, C.V.~Johnson and R.C.~Myers, \emph{{Charged AdS black
  holes and catastrophic holography}},
  \href{https://doi.org/10.1103/physrevd.60.064018}{\emph{Physical Review D}
  {\bfseries 60} (1999) }.

\bibitem{Taylor_ct}
M.~Taylor, \emph{{More on counterterms in the gravitational action and
  anomalies}},  \href{https://arxiv.org/abs/hep-th/0002125}{{\ttfamily
  hep-th/0002125}}.

\bibitem{Skenderis_notes}
K.~Skenderis, \emph{{Lecture notes on holographic renormalization}},
  \href{https://doi.org/10.1088/0264-9381/19/22/306}{\emph{Class. Quant. Grav.}
  {\bfseries 19} (2002) 5849}
  [\href{https://arxiv.org/abs/hep-th/0209067}{{\ttfamily hep-th/0209067}}].

\bibitem{weyl_anomaly}
M.~Henningson and K.~Skenderis, \emph{{The holographic Weyl anomaly}},
  \href{https://doi.org/10.1088/1126-6708/1998/07/023}{\emph{Journal of High
  Energy Physics} {\bfseries 1998} (1998) 023–023}.

\bibitem{Hsiao_Son}
W.-H.~Hsiao and D.T.~Son, \emph{Duality and universal transport in
  mixed-dimension electrodynamics},
  \href{https://doi.org/10.1103/physrevb.96.075127}{\emph{Physical Review B}
  {\bfseries 96} (2017) }.

\bibitem{Witten_multitrace}
E.~Witten, \emph{{Multitrace operators, boundary conditions, and AdS / CFT
  correspondence}},  \href{https://arxiv.org/abs/hep-th/0112258}{{\ttfamily
  hep-th/0112258}}.

\bibitem{Kim_num}
K.-Y.~Kim, K.K.~Kim, Y.~Seo and S.-J.~Sin, \emph{Coherent/incoherent metal
  transition in a holographic model},
  \href{https://doi.org/10.1007/jhep12(2014)170}{\emph{Journal of High Energy
  Physics} {\bfseries 2014} (2014) }.

\bibitem{Qlattices}
A.~Donos and J.P.~Gauntlett, \emph{Holographic q-lattices},
  \href{https://doi.org/10.1007/jhep04(2014)040}{\emph{Journal of High Energy
  Physics} {\bfseries 2014} (2014) }.

\bibitem{Grozdanov_GB}
S.~Grozdanov and A.O.~Starinets, \emph{{Second-order transport, quasinormal
  modes and zero-viscosity limit in the Gauss-Bonnet holographic fluid}},
  \href{https://doi.org/10.1007/jhep03(2017)166}{\emph{Journal of High Energy
  Physics} {\bfseries 2017} (2017) }.

\bibitem{Folkestad_2019}
A.~Folkestad, S.~Grozdanov, K.~Rajagopal and W.~van~der Schee, \emph{Coupling
  constant corrections in a holographic model of heavy ion collisions with
  nonzero baryon number density},
  \href{https://doi.org/10.1007/jhep12(2019)093}{\emph{Journal of High Energy
  Physics} {\bfseries 2019} (2019) }.

\bibitem{DHK_cMB}
E.~D’Hoker and P.~Kraus, \emph{{Charged magnetic brane solutions in AdS5 and
  the fate of the third law of thermodynamics}},
  \href{https://doi.org/10.1007/jhep03(2010)095}{\emph{Journal of High Energy
  Physics} {\bfseries 2010} (2010) }.

\bibitem{Vardhan:2022wxz}
S.~Vardhan, S.~Grozdanov, S.~Leutheusser and H.~Liu, \emph{{A new formulation
  of strong-field magnetohydrodynamics for neutron stars}},
  \href{https://arxiv.org/abs/2207.01636}{{\ttfamily 2207.01636}}.

\bibitem{Vardhan:2024qdi}
S.~Vardhan, S.~Grozdanov, S.~Leutheusser and H.~Liu, \emph{{Effective field
  theories of dissipative fluids with one-form symmetries}},
  \href{https://arxiv.org/abs/2408.12868}{{\ttfamily 2408.12868}}.

\bibitem{instab_plasma1}
F.F.~Cap, \emph{Handbook on Plasma Instabilities}, Elsevier Science \&
  Technology, United States (1982).

\bibitem{mikhailovskii2017instabilities}
A.B.~Mikhailovskii, \emph{Instabilities in a confined plasma}, CRC Press
  (2017).

\bibitem{Mtheory}
C.P.~Herzog, P.~Kovtun, S.~Sachdev and D.T.~Son, \emph{{Quantum critical
  transport, duality, and M-theory}},
  \href{https://doi.org/10.1103/PhysRevD.75.085020}{\emph{Phys. Rev. D}
  {\bfseries 75} (2007) 085020}
  [\href{https://arxiv.org/abs/hep-th/0701036}{{\ttfamily hep-th/0701036}}].

\bibitem{Giombi_confCFT}
S.~Giombi, G.~Tarnopolsky and I.R.~Klebanov, \emph{{On $C_{J}$ and $C_{T}$ in
  Conformal QED}}, \href{https://doi.org/10.1007/JHEP08(2016)156}{\emph{JHEP}
  {\bfseries 08} (2016) 156}
  [\href{https://arxiv.org/abs/1602.01076}{{\ttfamily 1602.01076}}].

\end{thebibliography}\endgroup

%% or
%% [B] Manual formatting (see below)
%% (i) We suggest to always provide author, title and journal data or doi:
%% in short all the informations that clearly identify a document.
%% (ii) please avoid comments such as "For a review'', "For some examples",
%% "and references therein" or move them in the text. In general, please leave only references in the bibliography and move all
%% accessory text in footnotes.
%% (iii) Also, please have only one work for each \bibitem.

\end{document}